\documentstyle[epsfig]{article}
\begin{document}
\begin{center}
\bf MULTIPION COHERENT EFFECTS IN HIGH ENERGY HEAVY-ION COLLISIONS\\
\vspace {1cm}
\rm G.Z.OBRANT and M.G.RYSKIN\\
\vspace {0.5cm}
\it Petersburg Nuclear Physics Institute, Gatchina, Leningrad
district, 188350 Russia
\end{center}
\begin{abstract}
Multipion production in high energy nucleus-nucleus collisions
is considered in the model of the pion 
radiation by classical current.   
Strong coherent effects  
of narrowing of pion longitudinal
and transverse momenta distributions
are predicted at the RHIC energy.
The coherence  enhances the large pseudorapidities 
producing the bump in the distribution.
The growth of the average pion multiplicity
and the oscillation effect in the multiplicity distribution are 
caused by the coherence as well.
\end{abstract}

\section {Introduction}

The observation of a new physical phenomenon could be done
in experiments at RHIC. 
It is a coherent multipion radiation in heavy-ion collisions 
analogous to the laser radiation [1,2] of the electromagnetic field.
Besides of it's proper interest this is  important now both in  
connection with DCC [3] and classical pion field [4] problems. 
It is known that quantum statistical correlations change the
multiplicity distribution and the momentum spectra of identical 
hadrons. The effect becomes crucial in a high-energy heavy-ion
collisions where the mean number of pions in a unit volume of 
phase space ($\Delta x\Delta k $ cell) reaches the value of about 
a unity (or more than a unity). In this case the coherent radiation of  
pions may lead to the formation of a so-called pion laser [1]. Usually 
these effects are discussed assuming a thermal emission of pions by 
a static source with radius $R $ [5].

On the other hand the most popular basic model to describe the
underlying events in a heavy-ion collision (assuming no any new 
physics in a nucleus-nucleus scattering) is a model of independent
nucleon-nucleon interactions. Each pair of incoming nucleons
produces secondary pions independently. In other words, each pair
of colliding nucleons plays the role of a separate source placed at
some point $x_i $ within the space-time domain where the beam $A_1 $ 
and target $A_2 $ ions overlap. Each source is more or less weak, 
still the number of sources are as large ($\sim A $) that 
the density of pions becomes greater than a unity (in a cell of 
phase space).
In the present paper we will study how the quantum statistics (i.e.
the permutations) of identical pions modifies the spectra and
multiplicity distribution of secondaries in such a simple (basic) 
model of independent nucleon-nucleon interactions
\footnote{This model may be considered as a realistic one if indeed 
there are no new dynamic effects in a nucleus-nucleus collision}.  

Some possible effects of the pion coherence  
was
considered qualitatively in the simple model of bremsstrahlung 
of scalar pions with a small pion multiplicity [6]. 
We do now the next step toward the quantitative understanding of this
phenomenon considering more realistic central nucleus-nucleus   
collisions and accounting true
pion amplitudes with the large multiplicity in the nucleon-nucleon 
interaction. This could be carried out using the model 
of Gyulassy, Kauffmann and Wilson [7] of the pion
radiation by classical current in the nucleus-nucleus scattering.
We see the strong coherent  
narrowing in the pion transverse 
and longitudinal momenta distribution for such a process. 
This confirms the analogous result obtained in [6].
The longitudinal momenta are
soften strongly by coherence contrary to the model of the scalar pions 
bremsstrahlung where the monochromatization of pions takes place.
In spite of this difference the conclusion [6] about the 
enhancement of large pseudorapidities by coherence is 
encouragingly confirmed now.
Besides the kinematical variables we have 
prominent coherent effects in the pion multiplicity
taking the large number of radiating nucleons.
That is in accordance with
the threshold character of the 
coherence versus the number of radiating nucleons [6].  

\section {The pion radiation by classical current}

The average pion multiplicity in the central heavy-ion collisions at  
RHIC energy will be of an order $10^4 $. The convenient way to do the 
calculation with such a number of radiated pions is to consider 
the pion radiation by classical current [7]. The S-matrix for the
emittance of m pions 
by the current $J({\bf k}) $ can be written as
\begin{eqnarray}
|J>=e^{-\bar n/2}\exp (i\int d^3kJ({\bf k})a^+({\bf k}))|0> \\
\bar n=\int d^3k|J({\bf k})|^2\qquad \qquad \qquad  \\
S_{fi}=<a({\bf k}_1)\cdots a({\bf k}_m)|J> \qquad \quad  
\end{eqnarray}
The coherent state $|J> $ is the state with the indefinite number of
quanta and the quantum-mechanical average of the 
pion annihilation operator 
$a({\bf k}) $ equals to $J({\bf k}) $. The exclusive cross section,
the semiinclusive cross section for the case of m radiated pions, the
inclusive cross section and the pion multiplicity distribution could
be easily obtained from Eqs.(1)-(3) using the permutation property
of operators $a({\bf k}_i) $ and $a^+({\bf k}_i) $.
\begin{eqnarray}
\frac {dW^{(m)}}{d{\bf k}_1\cdots d{\bf k}_m}=|J({\bf k}_1)|^2\cdots 
|J({\bf k}_m)|^2e^{-\bar n} \\
\frac {dW^{(m)}}{d{\bf k}_1}=
|J({\bf k}_1)|^2
\frac {\bar n^{m-1}}{(m-1)!}e^{-\bar n}\qquad \\ 
\frac {dW}{d{\bf k}_1}=\sum _{m=1}^{\infty }\frac 
{dW^{(m)}}{d{\bf k}_1}=|J({\bf k}_1)|^2\qquad \\ 
P_m=\frac {\bar n^m}{m!}e^{-\bar n}\qquad \qquad \qquad \qquad \quad    
\end{eqnarray}
So, we obtain the Poisson multiplicity distribution in this model, and
the physical sense of the value $\bar n $ (Eq.(2)) is clear from Eq.(7)
as the average pion multiplicity.

One can write the nuclear current as a sum of currents of the constituent
nucleons [7].
\begin{eqnarray}
J(x)=\sum _{i=1}^NJ_\pi (x-x_i),\quad J({\bf k})=
J_0({\bf k})\sum _{i=1}^Ne^{i\omega t_i-i{\bf k}{\bf x}_i}\qquad  
\end{eqnarray}
Here the space-time points $x_i $ are the coordinates of some "inelastic
scattering centers" where the strength of the current is localized. 
We obtain all distributions by formulas Eqs.(4)-(7) substituting
the following expression for $\bar n $ which depends now on the 
instant space-time coordinates of N nucleons $x_1,...,x_N $.
\begin{eqnarray}
\bar n=\bar n(x_1,...,x_N)=\int d^3k|J({\bf k})|^2
(N+2\sum _{i<j}\cos k(x_i-x_j)) 
\end{eqnarray}
The final cross section can be obtained by averaging 
of the distributions Eqs.(4-7) over the whole
space-time region of the nucleus-nucleus collision.

The amplitude  of the vacuum-vacuum transition   
$exp (-\bar n(x_1,...,x_N)/2) $ gives the right
cross section normalization, i.e., provides the unitarity 
being carried out. This takes into account the radiation of the pion 
in one point and its absorption in another one. 

As an example we study the central Au-Au 
collisions with the energy $p_0=100 GeV $ per 
nucleon in c.m.- system. The number of radiating nucleons N=150 is 
fixed. We neglect now spin and isospin variables
considering creation of pions only of one sign. The current 
$J({\bf k}) $ is obtained by fitting the single-inclusive pion 
longitudinal [8] and transverse 
momenta distributions 
with the real average pion 
multiplicity in a nucleon-nucleon scattering 
\footnote{We are grateful to Yu.M.Shabelsky for providing us
with tabulated pion distributions}. 
To average the cross section
over the radiation space-time region we approximate this domain by the
Woods-Suxon distribution with the radius $R=6.3fm $ and the diffuse 
parameter $a=0.6fm $. The longitudinal coordinate  
and the time of the radiation was averaged independently, the Lorentz 
contraction of the nucleus being taken into account.

The resulting distributions on the kinematical variables are shown
in Fig.1. The coherence is seen to produce strong effects both
in the transverse momentum distribution and in the longitudinal one.
Pion transverse
momentum is diminished by coherence to the value about of
the nucleus inverse radius. To estimate the characteristic pion energy
with this transverse momentum
lets take the corresponding magnitude  
in the maximum of the pseudorapidity distribution.
This energy $\rm \omega \approx  0.75 GeV $  
is somewhat lower than the inverse nuclear range shortened by the 
Lorentz boost. It corresponds to the fact that these pions are  
radiated from some more wide longitudinal region determined mostly by
the longitudinal size of the slow particle space in the nucleon.

The coherent decrease is more strong for pion transverse momenta than
that for longitudinal ones. As both $k_l $ and $k_t $ tend to the
corresponding inverse 
\newpage 
\vspace*{3.0cm}
\begin{figure}[h]
\vspace {-4.0cm}
\mbox{\hspace*{0.0cm}\epsfig{figure=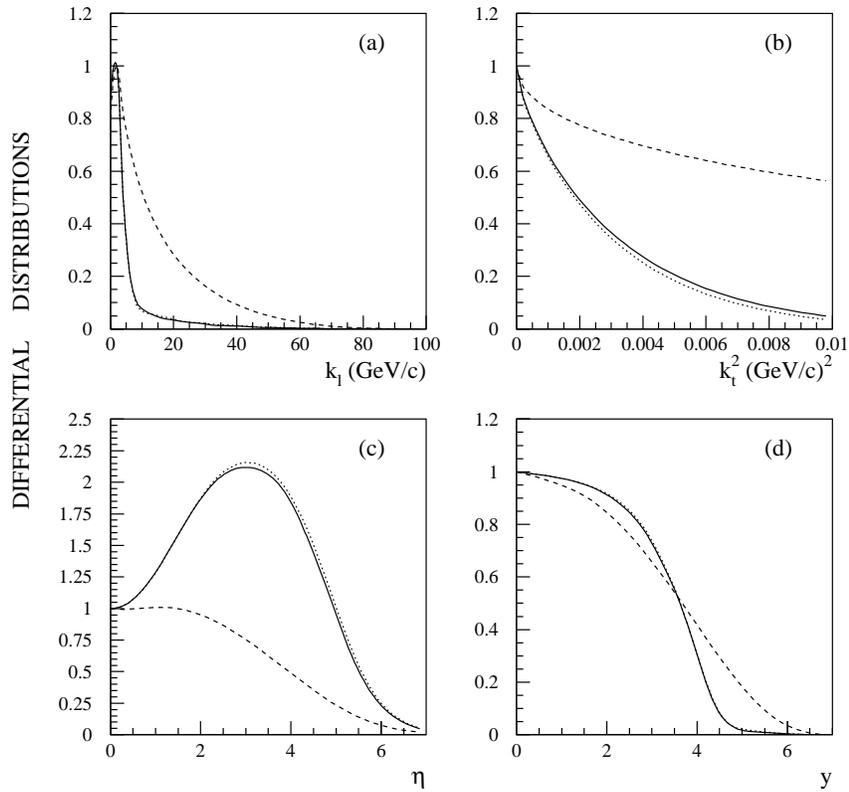,width=12cm}}
\caption{
Distributions of the pion longitudinal momentum
(multiplied by $k_l $)(a), transverse momentum square (b), 
pseudorapidity (c) and rapidity (d). The solid line -
the inclusive cross section
with the fixed number of pions in the event m=7000,
the dotted line - the inclusive cross section, the dashed line -
the inclusive cross section without the interference contribution 
}
\end{figure}
\newpage
\noindent
nuclear size, one may estimate the effect
as follows:
\begin{eqnarray}
k_l/k_{l0} \sim \frac {p_0}{m}\frac {1}{R}/<x_{F0}>p_0 \approx 0.3 ,
\qquad  
k_t/k_{t0} \sim \frac {1}{R}/<k_{t0}> \approx 0.1 \nonumber
\end{eqnarray}
The values $k_{l0} $ and $k_{t0} $ are characteristic pion
longitudinal and transverse momenta without the coherence,
$<x_{F0}> \approx 0.1 ,~<k_{t0}> \approx 0.36 GeV $ - 
the average Feynman's x and pion transverse
momentum in the nucleon-nucleon interaction and $m $ is the nucleon mass.
So, the large pseudorapidities must be enhanced by coherence. This 
effect is really seen in Fig.1(c),
the cross section being twice larger than 
the noncoherent one in the shoulder region having the bump here.
One could suppose
the behavior of this effect with the growth of $p_0 $ 
to be determined by the competition of two factors. 
Decrease of $<x_{F0}> $ and
increase of $<k_{t0}> $ must
enhance the effect. On another hand
the fact that the longitudinal coherent region tends to the 
hadron size $r_0 $ appeared to diminish the effect. 

The rapidity distribution (Fig.1(d)) shows coherent effect as well.
However, this is not convenient kinematical variable now to clear up
the physics when pion transverse momenta are lower than the pion mass. 
All considered inclusive distributions do not distinguish remarkably from 
the spectra where the number of pions is fixed. 

\section {Multiplicity distribution}

The existence of strong coherent effects in the pion multiplicity  
is suggested by the fact that the laser-type pion radiation has the
characteristic $N^2 $ dependence of the average multiplicity on the
number of nucleon-sources [7]. The averaged over the space-time region
value $\bar n ($ Eq.(9)) is given by
\begin{eqnarray}
<\bar n>=\bar n_0(N+\epsilon N(N-1))
\end{eqnarray}
Here $\bar n_0 \approx 6 $ is the average  
multiplicity of one sign pions in the nucleon-nucleon
interaction and $\epsilon $ is the probability for two pions to be in
the same state (i.e. in the same $ \Delta x\Delta k $ cell). 
The strong threshold growth of coherent effects in the
pion induced radiation versus the number of radiated nucleons [6]
corresponds to the small overlapping of the pion distributions
after the averaging  according to the first and the second term 
in Eq.(10).

We have calculated the pion multiplicity distribution Eq.(7) and  
have shown this in Fig.2. 
The average multiplicity is seen to be approximately
seven time larger than that in the noncoherent case. 
So, the second term dominates in Eq.(10) and we have the coherent 
source with thousands of pions in the same state. It is really a pion 
classical field. The probability $\epsilon \approx 0.04 $ could 
be obtained using Eq.(10). 

The prominent distinction of the obtained multiplicity distribution 
from the Poisson distribution, its large wide and
the oscillating shape 
is the second effect of the pion coherence. 
To understand the nature of this oscillations let's consider
all the process as a sum of  
coherent radiations by groups of nucleons, the nucleons in the  
group being distributed over the whole nucleus. 
Fitting the places of two maximums 
in the spectrum  by the expressions 
$\bar n_0\epsilon N_1(N_1-1))B $ and $\bar n_0\epsilon N_1(N_1+1))B $,
correspondent to groups  
with the number of nucleons 
distinguishing to a unity 
we obtain the effective number of 
nucleons 
in the group $ N_1\approx 32 $. 
As the  radiation of different
groups of nucleons is considered
to be coherent the factor $B $ is 
given by $ N_g(N_g-1) $ ,where $N_g $ is the number of groups.
We obtain $N_g\approx 5 $
in accordance with the fact that
the total number of nucleons is $N=150 $.
 
The only known for us example of the
oscillating multiplicity distribution is such a spectrum in the
multireggeon cutting mechanism [9] where the peaks in the distribution 
correspond to the integer multiple of the single reggeon multiplicity.
The principal difference of this from the result being discussed 
is the fact that the laser type radiation
has the characteristic $N^2 $ dependence of the average multiplicity
on the number of nucleons in the group. Together with the fact that
the number of pions $\bar n_0 $ emitted by each pair of colliding
nucleons (each current $J({\bf k}) $ ) increases with energy 
($\bar n_0\gg 1 $) it provides the reason 
why contributions of different configurations are separated clear 
in the spectrum.

\section {Spin effects}

We have not considered influence of the spin of 
nucleons to the discussed coherent effects.
The multiple pion production at high energy seems not to have 
the significant dependence on the nucleon spin 
in the single-inclusive pion spectra, at least in the central rapidity   
region or in the multiperipheral picture of a pion
creation. However, the feature of a pion being a Goldstone particle
to couple with spin (for ex. with the spin of a quark in the nucleon)  
suppressing 
\newpage
\vspace*{3.0cm} 
\begin{figure}[h]
\vspace {-4.0cm}
\mbox{\hspace*{0.0cm}\epsfig{figure=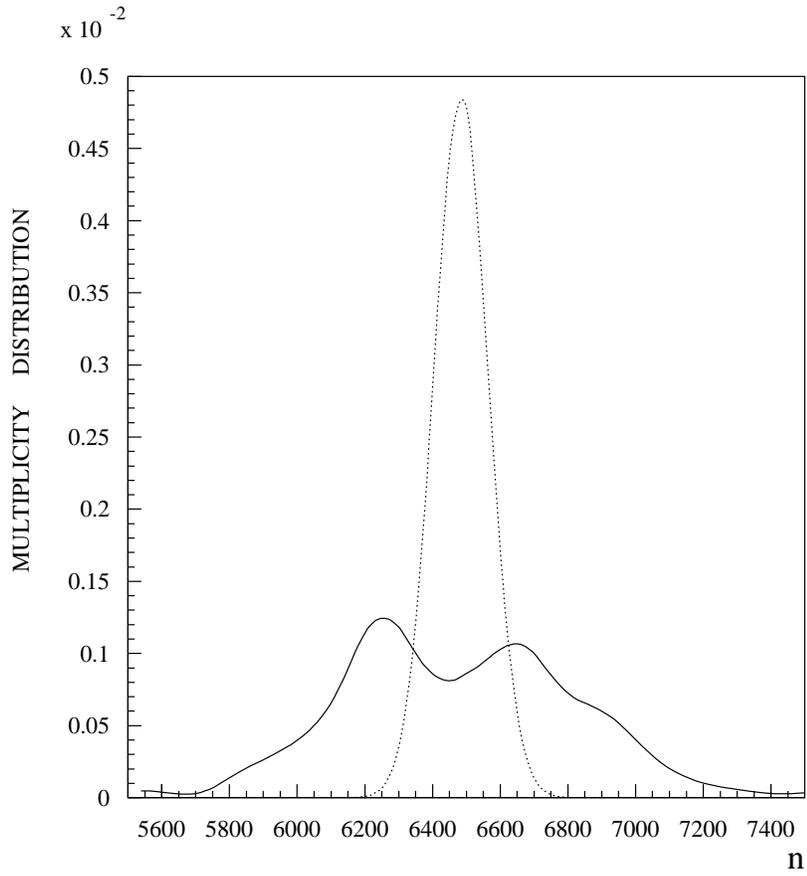,width=12cm}}
\caption{
The pion multiplicity distribution (the solid line) and the Poisson 
distribution with the same average multiplicity (the dotted line)
}
\end{figure}
\newpage
\noindent 
the small transverse momenta
demands to clarify the question about the spin dependence of the
discussed coherent effects
\footnote{We wish to thank B.L.Ioffe for paying our attention 
to this important point and for the interesting discussion}
.

First point is the fact that being a Goldstone the pion with zero
momentum $k $=0 (i.e. the homogeneous pion field) decouple from
quarks or barions. Next is the form of the pion-nucleon vertex;
in some sense pion is emitted by the spin of fermion. It is not
essential for us now either the fermion is a quark or a nucleon but
for simplicity bellow we will consider the nucleon-pion coupling.
In the nucleon rest frame the vertex takes the form 
$V=g({\bf \sigma }{\bf k}) $.

At first sight due to the Goldstone nature of pions one might expect 
the inclusive cross section to go to zero at small $k_t $, just in 
the region where we collect the main interference effects. However
this is not the fact: 
i) when the pion is emitted by a fermion, the vanishing of the vertex
$V \sim k $ is compensated by the pole of the fermion propagator and
at $k \to 0 $ the amplitude tends to a constant; 
ii) in a central rapidity region the majority (more than one half) of
pions comes from the resonance ($\rho ,\omega ,f,... $) decay. 
Anyway in our calculations we have used the experimental cross section 
(measured in $pp $-collisions) which does not go to zero as 
$k_t \to 0 $. 

Role of the spin is a more delicate question. First we have to
emphasize that in a central region the pions are produced mainly
due to the resonances decay and only small part of $\pi $-mesons are
created directly by fermions. Nevertheless, let us
discuss the interference between two identical pions (with momenta
$k_1 $ and $k_2 $) emitted by two nucleons with coordinates $x_1 $ 
and $x_2 $. The amplitude reads:
\begin{eqnarray}
A=g[({\bf \sigma }_1{\bf k}_1)e^{ik_lx_1}
({\bf \sigma }_2{\bf k}_2)e^{ik_2x_2}+
({\bf \sigma }_1{\bf k}_2)e^{ik_2x_1}
({\bf \sigma }_2{\bf k}_1)e^{ik_1x_2}]
\end{eqnarray}
where we keep only the terms (and factors) essential for our discussion.
To calculate the cross section one has to square the amplitude Eq.(11)
and to average over the nucleon polarizations. It leads to: 
\begin{eqnarray}
\frac {dW}{d{\bf k}_1d{\bf k}_2}\sim g^2[{\bf k}_1^2{\bf k}_2^2+
2({\bf k}_1{\bf k}_2)^2e^{i(k_1-k_2)(x_1-x_2)}+{\bf k}_2^2{\bf k}_1^2]
\end{eqnarray}
Here the second term corresponds to the interference of the amplitudes,
where the pions $k_1 $ and $k_2 $ are emitted by nucleons 1 and 2 
correspondingly, and vise versa. After the averaging over polarizations 
the product $({\bf \sigma _1}{\bf k}_1)({\bf \sigma _1}{\bf k}_2) $
gives the value $({\bf k}_1{\bf k}_2) $. Thus in comparison with a 
spinless case $dW\sim [1+\exp (i(k_1-k_2)(x_1-x_2))] $ we obtain an 
extra cosine square $({\bf k}_1{\bf k}_2)^2/{\bf k}_1^2{\bf k}_2^2=
\cos ^2\theta $. Accounting for the spin we get
$dW\sim [1+\cos ^2\theta \exp (i(k_1-k_2)(x_1-x_2))] $. 
Note that for ${\bf k}_1={\bf k}_2 $ the interference is still as 
strong as before. The only role of this $\cos \theta $ is a tiny 
decreasing of an effective volume of elementary cell 
($\Delta x\Delta k $) where the interference do takes place. In other
words taking the fermion spin into account we diminished a little bit
the value of $\epsilon $. It should be stressed that in the region
of interest $|\eta |\sim 3 $ the pion momenta $k_1 $,$k_2 $ are not
too small. In the nucleon rest frame we deal with the values of
$|{\bf k}_1| $,$|{\bf k}_2| \sim 1GeV $. On the other hand within 
the interference peak the difference $\Delta k=|{\bf k}_1-{\bf k}_2| $
is of the order of $1/R \approx 30MeV $. So the typical values of
$\cos ^2\theta \approx 1-(\Delta k)^2/k^2\approx 0.998 $ are very
close to 1.

To demonstrate the role of spin we consider numerically the production
of two pions emitted by a couple of nucleons in the 
deuteron-deuteron collision [6], the $\pi NN $ vertex being taken as
$\gamma _5 $. 
In our model pions are created 
by bremsstrahlung in the graphs Fig.3 of the old
perturbation theory.
\vspace {1.0cm}
\begin{figure}[h]
\mbox{\hspace*{3.0cm}\epsfig{figure=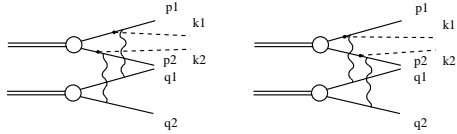
,width=6cm}}
\caption{
The amplitude of the pion radiation. 
}
\end{figure}
\vspace {0.5cm}
\noindent
Here the Goldstone nature of the bozon
should reveal itself as much as possible.
An one-loop amplitude is 
given by a threefold integral on the 
nucleon momentum in the intermediate state
\begin{eqnarray}
M=\int \frac {d^3p_1'}{(2\pi )^3}
\frac {1}{2E(p_1')2E(p_2')2E(q_1')2E(q_2')}
\frac {G(S_p)}{E_0-E(p_1')-E(p_2')}\times \nonumber \\
\times \frac {G(S_q)}{E_0-E(q_1')-E(q_2')}
\frac {1}{E_1+i\eta -E(p_1')-E(q_1')}
\frac {M_1}{h_1^2-m^2}\frac {M_2}{h_2^2-m^2} \\
h_i^2=(E(p_i)+E(q_i)-E(q_i'))^2-({\bf p}_i'-{\bf k}_i)^2\qquad 
(i=1,2).\nonumber
\end{eqnarray}
The value $E_0$ is the initial deuteron energy, 
$E(p)$,$\omega (k)$ - nucleon and meson energies 
and $E_1=E(p_1)+E(q_1)+\omega (k_1)$.
The formfactors
$G(S_p)$ and $G(S_q)$ with the corresponding 
energy denominators represent the 
wave functions of rapidly moving deuterons. 
The parameters of the simplest Hulthen 
wave-function have been chosen for the formfactor $G(S)$ 
producing the correct
nonrelativistic behavior in the two-nucleon rest-frame system.
The imaginary
part of the two-nucleon propagator dominates [6] with a good 
accuracy in Eq.(13).

The invariant amplitudes $M_1$ and $M_2$ contain amplitudes of the 
nucleon-nucleon elastic scattering multiplied by the 
spin-dependent part of the pion emission amplitude
\begin{eqnarray}
2m\bar u(p)(\hat p'-\hat k+m)\gamma _5u(p')S(k_l,k_t).
\end{eqnarray}
We include there the factor 
$S(k_l,k_t)=exp(-k_l/2k_{l0})exp(-k_t/k_{t0}) $
with $k_{l0}=0.15p_0 $ and $k_{t0}=0.36GeV $, providing 
reasonable agreement with experimental meson spectra.
The symmetrization of a sum of two time ordered
graphs Fig.3 was done for the
momenta ${\bf k}_1$ and ${\bf k}_2$ taking into account the pion 
identity. The deuteron was considered as a spinless particle and the 
resulting cross section was summed over final nucleon polarizations
and averaged over polarizations of nucleons in the deuteron. 
All integrations both internal in Eq.(13) and external
over the phase space was done using
the Quantum Mechanical Monte Carlo calculation [6].

The results of this calculation on pion distributions are shown
in Fig.4 for the interference part of the cross section. Such
interference contributions  was found to dominate in the cross  
section with the large number of pion sources [6].  
We see strong coherent effects both in the case with the account
of the nucleon spin (Fig.4(a,b)) and with its neglection 
(Fig.4(c,d)). 
The spin of nucleons in this
model demands of some more close momenta of two pions for the
interference to exist. As a result the narrowing of the longitudinal
and transverse momenta spectra is somewhat larger in this
case, the absolute value of the interference in
the maximum being approximately the same. 
We do observe some decrease of the interference contribution 
(of about 3.5$\% $) but it is too small 
to change our previous result noticeably.
\newpage 
\vspace*{3.0cm}
\begin{figure}[h]
\vspace {-4.0cm}
\mbox{\hspace*{0.0cm}\epsfig{figure=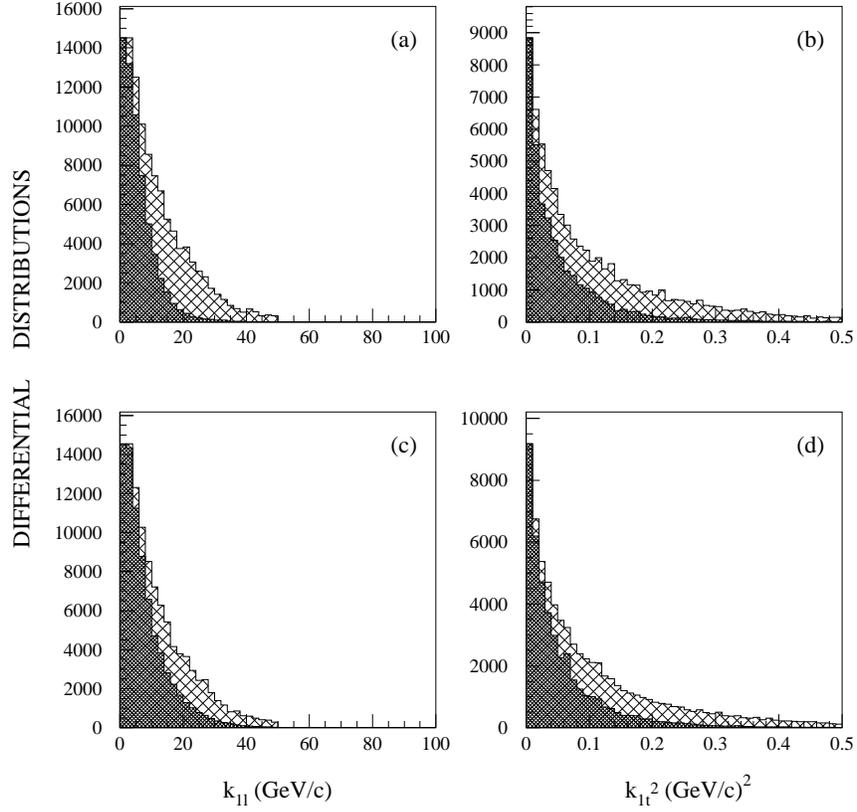,width=12cm}}
\caption{
The distributions of the pion longitudinal momentum
(a,c) and the transverse momentum square (b,d) in the case with 
the account of
the nucleon spin by Eq.(14) (a,b) or using the averaged square 
of Eq.(14) over nucleons polarizations (c,d). 
The dark histograms - the interference part of the cross section,
the light one - the noncoherent contribution. 
}
\end{figure}

\section {Comparison with data}

In spite of the fact that the model of a pion radiation by
classical current is rather crude and our results
have mostly qualitative character
we compare them with the existing experimental data.
We use results of two CERN experiments WA98 and NA49 on
Pb-Pb interaction with the energy 158A GeV [10]. 
The high centrality of the event was reached here
using the forward veto calorimeter
though this is somewhat lower than the value correspondent to N=150. 
The experimental efficiency falls in the low transverse  
momenta region $k_t < 0.2 GeV $ in both experiments.
This point is the most crucial for the observation of coherent 
phenomena due to the fact that these effects 
are mainly situated in the region
$k_t \sim 1/R $, i.e., at $k_t < 0.1 GeV $. We roughly approximated 
this inefficiency by the constant cross section at $k_t < 0.2 GeV $.

In Fig.5 we show the results of the calculation for the Pb-Pb 
interaction. The theoretical cross section is twice increased  
reflecting the crude character of our estimation of the experimental 
efficiency. We see 
the experimental data do not contradict to the discussed  
coherent effects both for the pion multiplicity
and 
for the shape of the rapidity distribution.
Note that the experimental spectrum in Fig.5(b) is even more narrow
than the theoretical one. 
This fact encourages us and we consider it as an argument in favour 
of an important role of coherence effects. 
The future experiments
measuring small pion transverse momenta could clear observe the pion
coherence.   

\section {Conclusions}

The nucleons are found to throw off their pion "fur-coats" 
coherently in the central nucleus-nucleus collisions 
very willingly. A lot of pions being in one state is created
in this radiation. 
The natural question one could inquire about is 
the interaction of pions during the short time when the nuclei 
overlapping  and the pions interaction in the final state.
The first kind of interactions seems to be not so important.
Recall the fact that the experimental pion transverse momentum 
distribution in heavy-ion collision is not remarkably wider than 
that in the nucleon-nucleon interaction.  
We can suggest the scenario of the further
evolution of the produced classical pion field as the selfinteracting
field [4,11]. The basis for such a hope is the smallness of pion
transverse momenta. The relative pion energy is small  
(of the order of the pion mass) in this case 
\newpage
\vspace*{3.0cm}
\begin{figure}[h]
\vspace {-4.0cm}
\mbox{\hspace*{0.0cm}\epsfig{figure=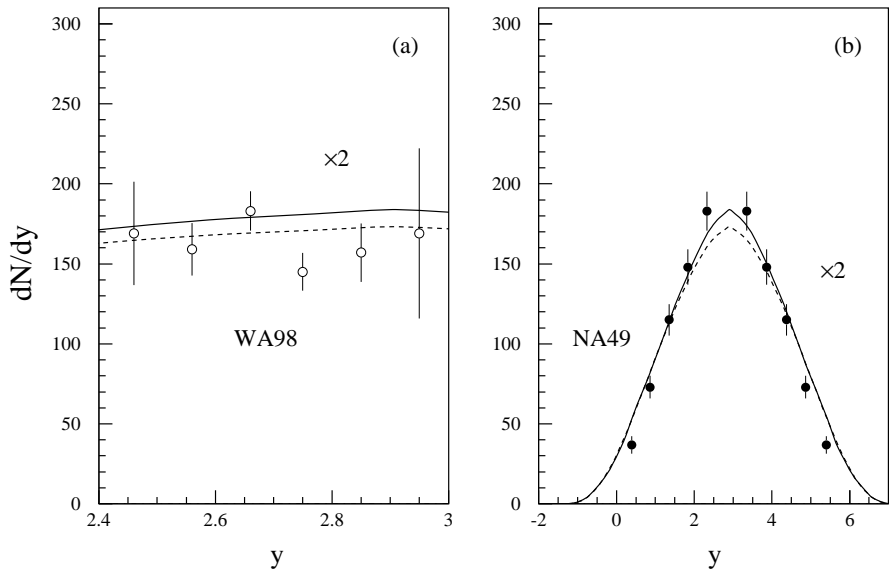,width=12cm}}
\caption{
Inclusive pion rapidity distributions (the average number of 
particles per event) with the interference contribution (the solid 
line) and without this (the dashed line).  
The data are for $\pi ^0 $ mesons (a) and for negative hadrons (b).
The theoretical values are multiplied by the factor two. 
}
\end{figure}
\noindent 
and one could use the soft pion theory [4]. 
The production of a classical field "brizer" [11] corresponding 
to the space-time region of the nucleus-nucleus collision 
could take place. When radiated, pions appeared to have some
characteristic longitudinal momenta in this case.
The usual final state interaction could take out from the coherent 
volume only small part of pions with large relative energies. 

We have to emphasize that at RHIC energies the effect of coherence 
essentially deforms the momentum spectra of secondary pions and
enlarges the multiplicity in comparison with the naive 
conventional estimates based on the model of independent
nucleon-nucleon interactions (without accounting for the coherence
of pions). The crucial point is the longitudinal Lorentz contraction 
of colliding nuclei. Therefore all the pions 
produced with $k_t\sim 1/R $ are emitted coherently.

The only way to reduce the effect shown in Fig.1,Fig.2 is to say 
that the pions are formed much after the collision. For example, 
if at first stage the Quark Gluon Plasma (QGP) would be created
and then (after the expansion and cooling of QGP) the pions would
be produced from the domain with the longitudinal size 
$\Delta z \sim 10 fm $, then the effect of coherence would be 
suppressed strongly.

So the absence of prominent coherent effects at RHIC may be 
considered as an argument in favour of QGP (or another new phase)
formation; the pions are emitted from a large size domain
after the "decay" of this new phase. Of course, in reality there
will be the competition between the direct production of a classical
(coherent) pion field 
and the formation of
another (like QGP) phase.

\end{document}